# Icosahedra boron chain and sheets: new boron allotropic structures

C. B. Kah, M. Yu[*], P. Tandy[†], C. S. Jayanthi, and S. Y. Wu

Department of Physics and Astronomy, University of Louisville, Louisville, Kentucky, 40292

ABSTRACT

The icosahedra boron chain and three icosahedra sheets (with α, $\delta_4$, and $\delta_6$ symmetries), constructed by the icosahedra $B_{12}$, have been obtained as new members of boron family using a highly efficient molecular dynamics scheme based on a transferable and reliable semi-empirical Hamiltonian. The icosahedral $B_{12}$ in the icosahedra chain is slightly elongated along the china direction and directly bonded each other with the two-center covalent bonds. A deformation of the icosahedra $B_{12}$ was also found in the two-dimensional icosahedra sheets. In addition to the three-center bonding nature inside the icosahedra $B_{12}$, there are two types of directional inter-icosahedra bonds in the icosahedra sheet structures, one is the single strong covalent bond, and the other is a pair of the weak covalent δ bonds. In contrast to the boron monolayer, there is no buckling found in these icosahedra sheets. The deformation of the icosahedra $B_{12}$ and the special bonding nature in these new icosahedra structures induce the energy band gap of 0.74 eV in the icosahedra chain, 0.52 eV in the icosahedra $\delta_6$ sheet, 0.39 eV in the icosahedra $\delta_4$ sheet, and the gapless in the icosahedra α sheet, respectively. The energy barrier per atom from the icosahedra $\delta_6$ sheet to the icosahedra α sheet is estimated to be 0.17 eV while it is estimated as 0.38 eV from the icosahedra $\delta_6$ sheet to the icosahedra $\delta_4$ sheet and 0.27 eV from the icosahedra α sheet to the icosahedra $\delta_4$ sheet, respectively. Such high energy barriers indicate that these icosahedra sheets are relatively stable.

[*]Email: m0yu0001@louisville.edu

[†]Current address: Defense Threat Reduction Agency, 8725 John J Kingman Rd, Stop 6201, Fort Belvoir VA, 22060

I. INTRODUCTION

Recently, researches on boron based materials have tremendously increased in the experimental synthesis and the computational modeling. The most interest on boron is due to the fact that it is a trivalent element (i.e., three valence electrons shearing four valence orbitals). Such electron-deficiency makes boron complicated in forming chemical bonds. It possesses the tendency to form either strong directional covalent bond or the three-center and two-electron bond. Therefore, analog to its neighbor element (carbon) in the periodic table, boron have various allotropic structures. The linear, planar, quasi-planar, convex, ring, and icosahedra structures are found in the small size of boron clusters [1-10]; the cage- and fullerene-like structures [11-17], as well as compact clusters [18-20] are postulated in the intermediate size of boron clusters. Meanwhile, the tubular and monolayer sheet structures have been predicted [21-26], and the stable pure boron crystalline structures including the rhombohedra $\alpha$-$B_{12}$, $\beta$-$B_{106}$, and $\gamma$-$B_{28}$ phases (referred as $\alpha$-B, $\beta$-B, and $\gamma$-B, respectively), as well as boron nanowires have recently been successfully synthesized [27-36]. Associated with such quick development in discovering boron allotropes, one of the remaining interest questions is: *are there any other allotropic structures of boron unexplored?* If yes, are the new allotropic structures interest and useful in the applications to the nanotechnology such as nanoelectronics, optics, and sensors? In this paper, we will take the challenge to answer these questions. We are interested in exploring the new nanostructures analogue to the boron crystalline structures. Our motivation is based on the fact that the boron crystalline structures, especially the $\alpha$-B in the ambient condition, are the most stable structures among boron allotropes [29] with various interesting chemical and physical properties, and structures analogue to the crystalline structures are expected to maintain the basic crystal properties.

To find such nanostructures, we considered the crystal structure of $\alpha$-B as the benchmark. The structure of the $\alpha$-B is build up with the icosahedra $B_{12}$ located at each vertex of the rohmbohedron, and each icosahedra $B_{12}$ is strongly and directly bonded to its neighbors making the $\alpha$-B boron the most stable. This special structure demonstrates that even though an isolated icosahedra $B_{12}$ is found to be energetically unstable compared to the energetically favored quasi-planar $B_{12}$ or the ring $B_{12}$ cluster, the aggregation of the icosahedra $B_{12}$ however, has shown to have the possibility to form stable polymorphs if such polymorphs are capable of forming the strong two-center covalent bonds between the icosahedra $B_{12}$. Stimulated by such considerations, we propose to construct the new boron allotropic structures using the icosahedra $B_{12}$ as the building block (referred as the icosahedra structures). Through a systematic molecular dynamics simulations we have discovered four interesting boron icosahedra structures, the icosahedra chain and three icosahedra sheets (referred as the icosahedra α, $δ_4$, and $δ_6$ sheets, respectively). These novel icosahedra structures are found to be structural and energetically stable and exhibit interesting chemical bonding nature and electronic properties. Especially, different from the α-B which has six strong two-center and twelve weak three-center inter-icosahedra bonds, the inter-icosahedra bonds in the icosahedra sheets are all directional two-center covalent bonds. Furthermore, the icosahedra $δ_6$ sheet, analogue to the surface structure along one of the rhombohedra face of the α-B, has been found to be more stable among the new icosahedra structures and is expected to be synthesized easily (e.g., cutting from the α-B). We have also found that theses icosahedra structures, except the gapless material of the icosahedra α sheet, exhibit semiconducting nature with the calculated energy band gap in the range of 0.39-0.74 eV, providing the possible applications for nanoelectronics. The modeling method that we employed in present work is based on a highly efficient semi-empirical Hamiltonian, referred as SCED-LCAO [18, 37, 38]. A detail description of our modeling using this approach will be given in section II. The results and discussions will be presented in details in section III. The possible pathways to synthesize these icosahedra sheet structures are also discussed in section III. Finally, the significant features of these novel icosahedra structures are summarized in the conclusion (section IV).

## II. MODELING

The semi-empirical SCED-LCAO Hamiltonian (i.e., the self-consistent and environment dependent semi-empirical Hamiltonian in the framework of the linear combination of the atomic orbitals) based molecular dynamics simulation scheme, developed by the condensed matter physics group at University of Louisville [37, 38], has been employed in this study. The detail description for the SCED-LCAO Hamiltonian and its various applications for B-, C-, Si- and Ge-based nano-materials have been reported in Refs [18, 37-44]. The most advantageous feature of the SCED-LCAO Hamiltonian, compared to other existing semi-empirical Hamiltonians, is that it has a framework to allow the self-consistent determination of charge-redistribution and the inclusion of the environment-dependent multi-center interactions on the same footing. These are two key ingredients for an appropriate description of bond-breaking and bond-forming processes that play dominant roles in the structure reconstruction of complex systems. Furthermore, an environment-dependent excitation orbital energy term has been included in the SCED-LCAO Hamiltonian to take into account the effect of the atom aggregation on the atomic orbital energy within the minimum orbital basis [18]. Such semi-empirical Hamiltonian for boron has shown its capability to characterize the complex chemical properties of the trivalent boron element and capture various bonding natures (e.g., the two-center and the three-center bonds) in the allotropes of boron including the small $B_n$ ($n$ = 2-24) clusters with the linear, planar, qausi-planar, convex, icosahedra and ring structures, the isotropic $B_{80}$ buckyballs with $I_h$, $T_h$, and $C_{2h}$ symmetries, the monolayer sheets with the buckled $\delta_6$, the buckled $\alpha$, and the flat $\delta_4$ symmetries, and the rhombohedra structure of the crystalline α-B (see details in Ref. [18]). It has also been applied to study the compact boron clusters $B_n$ ($100 < n < 800$) [18]. Such successful testimonies and applications demonstrate that the SCED-LCAO Hamiltonian for boron is transferable, reliable, and has the predict power. Therefore, in this work we will adopt this SCED-LCAO Hamiltonian to perform a comprehensive molecular dynamics study to explore the low dimensional boron icosahedra structures.

The low-dimensional icosahedra structures were constructed based on the structure of the α-B. As shown in Fig. 1(a), the structure of α-B has rhombohedra symmetry characterized with the lattice parameters of $a_0$ (the lattice constant) and α (the apex). Each icosahedra $B_{12}$ in the α-B has six covalent bonds (i.e., two-center inter-icosahedra bonds) bonded to atoms of its six nearest neighbor (NN) icosahedrons, indicated by $b_{inter}$ in Fig.1 (a). As founded from the experiments [32], the inter-icosahedra bonds $b_{inter}$ (~1.71 Å [32, 33]) are slightly shorter than the three-center bonds between boron atoms inside the icosahedra $B_{12}$ (referred as the intra-icosahedra bonds $b_{intra}$ in Fig.1 (a) which is in the range of 1.73 - 1.79 Å [32]). The interatomic distance (indicated by $d$ in Fig. 1 (a)) between the icosahedra $B_{12}$ and the atoms of its next nearest neighbor (NNN) icosahedra $B_{12}$ is longer than the two-center inter-icosahedra bond (~2.03 Å [32]). The one-dimensional icosahedra structure (referred as icosahedra chain) is then built up by a linear truncation of the α-B along one of its the lattice vectors (see Fig. 1 (b)) with the lattice constant of $a$. The two-dimensional icosahedra structures, on the other hand, are constructed with the icosahedra $B_{12}$ placed on the plan. Namely, the icosahedra $δ_6$ sheet has a triangular symmetry (see Fig. 1 (c)). It can be obtained by a plan truncation of the α-B along one of its rhombohedra faces and the lattice vectors of its primary unit cell are the two of the lattice vectors of the α-B along this face with the lattice constant of $a$. The icosahedra α sheet is then constructed by removing one icosahedra $B_{12}$ and creating one hexagonal hole from the 3x3 unit cells of the icosahedra triangular $δ_6$ sheet (see Fig. 1 (d)), and the icosahedra sheet $δ_4$ is constructed by removing one icosahedra $B_{12}$ and creating one hexagonal hole from the 2x2 unit cells of the icosahedra triangular $δ_6$ sheet (see Fig. 1 (e)). The notation of these icosahedra sheet symmetry follows the same notation for monolayer boron sheets defined in Ref [45]. Different from the icosahedra triangular $δ_6$ sheet, in the icosahedra α sheet there are three types of the icosahedra $B_{12}$ with different numbers of the single covalent inter-icosahedra bonds and the pairs of the inter-icosahedra δ bonds (referred as Type A, B, and C in Fig. 1 (d)). Type A of the icosahedra $B_{12}$ has four single inter-icosahedra bonds and two pairs of the δ bonds, similar as in the case of the icosahedra $δ_6$ sheet, type B of the icosahedra $B_{12}$ has three inter-icosahedra bonds and two pairs of the δ bonds, and type C of the icosahedra $B_{12}$ has four inter-icosahedra bonds and one pair of the δ bonds, respectively. In the icosahedra $δ_4$ sheet, on the other hand, there are two type of the icosahedra $B_{12}$ with different numbers of the single inter-icosahedra bonds and the pairs of the inter-icosahedra δ bonds (referred as Type A and B in Fig. 1(e)). Type A has only four single inter-icosahedra bonds, and type B has two single inter-icosahedra bonds and two pairs of the δ bonds, respectively. These low-dimensional icosahedra structures are then fully relaxed using the molecular dynamics version of the SCED-LCAO method. The time step in the molecular dynamics simulations was set to be 1.2 fs and the force criteria for a fully relaxation process was set to be less than $10^{-2}$ eV/Å.

In the meantime, we also employed the density functional theory (DFT) based VASP package [46] to further validate existence of the new low-dimensional icosahedra structures obtained from the SCED-LCAO method. The Vanderbilt Ultra-soft pseudo-potential [47, 48] was used to describe interaction between the core and the valence electrons, and the GGA PW91 version [49] of approximation was used for the exchange-correlation energy. To ensure the convergence of the total energy, the cut-off energy for the plane wave basis set was taken to be 321.5 eV, and several sets of *k* points were taken according to the Monkhorst-Pack scheme (*i.e.*, 1x1x6 for the icosahedra chain, and 5x1x5 for the icosahedra sheets, respectively). A vacuum region (15 Å) was chosen to ensure that there was no interaction between the periodic icosahedra chains or sheets. The structural relaxation in VASP calculations was performed using C.G. algorithm with the force criteria to be less than $10^{-3}$ eV/Å.

III.  RESULTS AND DISCUSSIONS

As a benchmark, we first calculated the structural properties of the α-B. The optimized lattice parameters for α-B from both the SCED-LCAO method [18] and the DFT method [46] are $a_0 = 5.058$ Å and $\alpha = 57.85^0$, in good consistent with the experimental measurements ($a_0 = 5.057 \pm 0.003$ Å and $\alpha = 58.06 \pm 0.05^0$) [32, 33]. The icosahedra chain and sheet structures are then optimized by scaling their lattice parameter $a$ to the optimized lattice constant $a_0$ and relaxing them for giving a value of $a$. Obtained relative energy as a function of the ratio of $a/a_0$ for these icosahedra structures is shown in Fig. 2 (a)-(d). The relative energy is defined to be the total energy per atom of the concerned system to that of the fully relaxed icosahedra $\delta_6$ sheet at its equilibrium. Apparently, the relative energy obtained from the SCED-LCAO method are consistent with that obtained from the DFT results (see the insets of Fig. 2 (a)-(d)) both in the shape of the curves and in the energy order among these icosahedra structures). Optimized lattice constant for the icosahedra chain (Fig.2 (a)) is 5.349 Å (5.297 Å in DFT) which is about 5.7 % larger than that of the α-B. On the other hand, the optimized lattice constants for the icosahedra α and $\delta_4$ sheets (Fig.2 (b) and (c)) are the same as 5.007 Å (5.058 Å in DFT), and the optimized lattice constant for the icosahedra $\delta_6$ sheet (Fig.2 (d)) is 4.957 Å (4.932 Å in DFT) which is about 1.9 % smaller that of the α-B.

Fig. 2 (a)-(d) shows that energetically, the icosahedra $\delta_6$ sheet is the lowest in energy among these icosahedra structures, followed by the icosahedra α sheet and the icosahedra $\delta_4$ sheet. The icosahedra chain is highest among them. This is because in the icosahedra $\delta_6$ sheet, each icosahedra $B_{12}$ is bonded to its six neighbors located at the vertex of a hexagon, four of them are bonded in the form of the single directional covalent bonds, and two of them are bonded with the pair of the δ bonds (see Fig. 1 (c)). Therefore, there are totally eight inter-icosahedra bonds. While, in the case of the icosahedra α sheet, even though the type A icosahedron (in Fig. 1 (d)) has the same bonding nature as that in the icosahedra $\delta_6$ sheet, the types B and C lose one icosahedra $B_{12}$ neighbor due to the existence of the hexagonal holes. The icosahedra $B_{12}$ near the hexagonal holes have either seven inter-icosahedra bonds (the type B in Fig. 1(d)) or six inter-icosahedra bonds (the type C in Fig.1 (d)). So, it is weaker in energy than the icosahedra $\delta_6$ sheet. In the case of the icosahedra $\delta_4$ sheet, each icosahedron has only four inter-icosahedra bonds (type A in Fig.1 (e)) or six inter-icosahedra bonds (type B in Fig. 1 (e)) and therefore, it is slightly higher in the energy than both of the icosahedra $\delta_6$ and α sheets. Similarly, there are only two inter-icosahedra bonds for each icosahedron $B_{12}$ in the icosahedra chain (see Fig. 1 (b)), and it has higher energy. We also investigated an icosahedra sheet with the hexagonal symmetry. In such system, there are only three neighboring icosahedra $B_{12}$ units per icosahedra $B_{12}$ unit with $120^0$ degrees. We found that such type of inter-icosahedra bonding does not fit the preference of the icosahedra in the crystalline structures and could not stabilize this icosahedra sheet. As the results, the icosahedra $B_{12}$ units become distorted and rotated to form a corrugated sheet with irregular orientation of the icosahedra $B_{12}$ and lead to the icosahedra sheet unstable.

The inter-icosahedra bonding nature in the obtained icosahedra sheets is interesting as compared to the bonding nature in the α-B. Based on the Longuet-Higgins & Roberts analysis [50], due to the electron deficiency in boron, the icosahedra $B_{12}$ has thirty six valence electrons but forty eight valence molecular orbitals. These orbitals include twelve outward-pointing radial orbitals and thirty six orbitals within the icosahedron with thirteen bonding and twenty three anti-bonding orbitals. A stable icosahedra $B_{12}$ is then need twenty six electrons to occupy the thirteen bonding orbitals and forming three-center two-electron type of bonds inside the icosahedron. The remaining ten electrons have to go the twelve outward-pointing orbitals and have the tendency to form bonds with the external neighboring atoms. When the icosahedra $B_{12}$ aggregate to form α-B, it is found that among the ten electrons, six electrons from six vertexes (atoms) of the icosahedron bond to the atoms of its six NN icosahedra $B_{12}$ units forming six directional covalent inter-icosahedra bonds, and the remaining four electrons from the other six vertexes (atoms) of the icosahedron participate the formation of the twelve weak three-center type of inter-icosahedra δ bonds with atoms of its six NNN icosahedra $B_{12}$ units. Such special and high symmetric bonding nature makes the α-B the most stable in the boron allotropes. When the icosahedra $B_{12}$ units form the two-dimensional structures such as the icosahedra $δ_6$ sheet, we found that (1) instead of six electrons participating in the formation of the six covalent inter-icosahedra bonds in the α-B, only four electrons (from four atoms of the icosahedra $B_{12}$) occupy the four outward-pointing orbitals and forming four single directional inter-icosahedra bonds with its four NN icosahedra $B_{12}$, namely, there are two dangling bonds at two vertex atoms; (2) instead of the four electrons participating the twelve three-center type of the inter-icosahedra δ bonds in the α-B, the four electrons (from other four atoms) occupy the other four outward-pointing orbitals forming the two-center type of the inter-icosahedra δ bonds to its two NN icosahedra $B_{12}$. Such δ bonds always appear in pairs, and their bond lengths are about 0.20 Å shorter than that in the α-B, indicating the strong interaction between those NN icosahedra $B_{12}$. Of course, there are two more dangling bonds at the remaining two atoms, totally four dangling bonds at the four vertexes of the icosahedron. Similarly, there are more dangling bonds in the icosahedra α and $δ_4$ sheets since there are less inter-icosahedra bonds, and therefore, less stable than the icosahedra $δ_6$ sheet. On the other hand, the icosahedra chain has only two electrons participating the two directional covalent inter-icosahedra bonds per icosahedron, and the remaining eight electrons stay in the remaining ten uncoupled orbitals, leading to its less stable than the icosahedra sheets.

The stability and the existence of these optimized low-dimensional icosahedra structures are further confirmed by the calculation of their lattice vibration frequencies at their equilibrium. The vibration frequency at gamma point, obtained by solving the eigenvalue problem of the force constants in VASP, ranges from 216.640 cm$^{-1}$ to 1136.765 cm$^{-1}$ for the icosahedra $\delta_6$ sheet, from 38.116 cm-1 to 1159.927 cm-1 for the icosahedra α sheet, from 34.416 cm$^{-1}$ to 1197.964 cm$^{-1}$ for the icosahedra $\delta_4$ sheet, and from 43.599 cm-$^{1}$ and 1187.738 cm$^{-1}$ for the icosahedra chain, respectively (see the frequency density in Fig. 3 (a)-(d)). The positive vibration frequencies indicate that even though these low-dimensional icosahedra structures are energetically higher than the α-B (e.g., 0.28 eV/atom for the icosahedra $\delta_6$ sheet), they could be exist or synthesized under some special condition, such as the high pressure or particular means (e.g. etching α-B or laser ablation). Recent experimental report on the growth of the crystalline boron nanowires by chemical vapor deposition was found to have the orthorhombic symmetry [34], providing the possible way to synthesize the icosahedra structures.

The structure properties of these new icosahedra structures were analyzed by examining their local structure (see Table 1), the pair-distribution functions (Fig. 4 (a) and Fig. 5(a)), and the angle-distribution functions (Fig. 4 (b) and Fig. 5 (b)), respectively. Table 1 lists (1) the bond length between boron atoms inside the icosahedra B$_{12}$, $b_{intra}$, (2) the bond length of the inter-icosahedra bond, $b_{inter}$, where the notation S is for the single covalent inter-icosahedra bond, and D, for the pair of the inter-icosahedra δ bonds, respectively, (3) the number of the covalent inter-icosahedra bonds per icosahedra B$_{12}$, $N_{inter}^{S}$, and (4) the number of the pairs of the inter-icosahedra δ bonds per icosahedra B$_{12}$, $N_{inter}^{D}$, respectively. Note that the total number of the inter-bonds per icosahedra B$_{12}$ should be the sum of the $N_{inter}^{S}$ and $N_{inter}^{D}$ (i.e., $N_{inter}^{Total} = N_{inter}^{S} + 2N_{inter}^{D}$).

It can be seen from the Table 1 that the intra-bond $b_{intra}$ in the icosahedra chain (i.e., 1.62-1.75 Å) is similar to that of an isolated icosahedra B$_{12}$ (i.e., 1.61-1.75 Å), but smaller than those of α-B (i.e., 1.73-1.80 Å). The covalent inter-icosahedra bonds $b_{inter}$ (i.e., 1.76 Å), on the other hand, is about 0.9 Å larger than that of the α-B (i.e., 1.67 Å from both the SCED-LCAO and DFT results), indicating the weak interactions between the icosahedra $B_{12}$ units in the icosahedra chain. The first peak in the pair distribution functions of the icosahedra chain (black solid curve in Fig. 4 (a)) has similar shape to that of the isolated icosahedra $B_{12}$ (the black dashed curve in Fig. 4 (a)). The shoulder of the first peak around 1.76 Å mostly represents the inter-icosahedra bond length. The periodicity of the icosahedra chain is characterized by the pronounced peaks at long distances. The peak around $60^0$ degree (describing the triangular nature inside the icosahedra $B_{12}$) in the angle-distribution function of the icosahedra chain (black solid curve in Fig. 4 (b)) is split by several distinguished sub-peaks as comparing with the single peak around $60^0$ degree for icosahedra $B_{12}$ (black dashed curve in Fig. 4 (b)). Such splitting corresponds to the deformation or symmetry reduction along the chain axis when the icosahedra $B_{12}$ form the chain structure. For instance, the peak around $96^0$ and $128^0$ degrees come from the elongation of the icosahedra $B_{12}$ unit along the chain axis (see black dotted lines inside the structure of the icosahedra chain in the inset of Fig. 4 (b)).

Different from the icosahedra chain, there are several interesting structural properties found in the icosahedra sheets. First, the distribution of the intra-bond length $b_{intra}$ is much wider in the icosahedra sheets (e.g., 1.56-1.77 Å in the icosahedra $\delta_6$ sheet, 1.56-1.80 Å in the icosahedra α sheet, and 1.59-1.80 Å in the icosahedra $\delta_4$ sheet, respectively) than in the α-B (i.e., 1.73-1.80 Å), indicating a slightly deformation of the icosahedra $B_{12}$. This phenomenon can also be seen from the broadening of the first peak in the pair-distribution functions of the icosahedra sheets (see the black solid, the red dashed, and the green dotted-dashed curves in Fig. 5 (a)) with respect to that of the α-B (see the black dotted curve in Fig. 5 (a)). Second, the single covalent inter-icosahedra bond length $b_{inter}$ (S) is similar to that of the α-B (e.g., 1.69 Å for the icosahedra $\delta_6$ sheet, 1.69 Å (Type A) or 1.69-1.72 Å (Type B and C) for the icosahedra α sheet, and 1.68 Å (Type A and B) for the icosahedra $\delta_4$ sheet, respectively), indicating that the strength of the interaction between these such icosahedrons is the same as in the α-B. Third, the paired inter-icosahedra δ bonds, indicated by (D) in the third column of Table 1, are much shortened in the case of the icosahedra sheets (e.g., 1.80 Å in the icosahedra $\delta_6$ sheet, 1.82 Å (Type A and B) and 1.75 (Type C) in the icosahedra α sheet, and 1.82 Å (Type B) in the icosahedra $\delta_4$ sheet, respectively) as compared to that in the α-B (i.e., 2.02 Å), indicating the strong interactions between these icosahedrons. That is why the sub-peak round 2.0 Å in the pair-distribution function of the α-B disappears in the case of the icosahedra sheets (see the black dotted curve in Fig. 4 (a)). Furthermore, the reason why $b_{inter}$ (D) in the icosahedra α sheet is shorter in the type C than in the types A and B is that the icosahedra $B_{12}$ located at the type C position has only one pair of the δ bonds due to the existence of the hexagonal hole. Forth, for each icosahedra $B_{12}$, its number of the single covalent inter-icosahedra bonds ($N^S_{inter}$, the 4$^{th}$ column in Table 1) and the pair of the inter-icosahedra δ bonds ($N^D_{inter}$, the 5th column in Table 1) depends on the environment of the icosahedra $B_{12}$ in the sheet, in particular, in the cases of the icosahedra α and $\delta_4$ sheets. The above analyses clearly demonstrate that the slight deformation of the icosahedra $B_{12}$, the covalent bonding nature in both the single and pair inter-icosahedra bonds, and the shortness of the inter-icosahedra δ bonds in the icosahedra sheets play crucial roles in stabilizing the icosahedra sheets. Such symmetry reduction induced bonding nature in the icosahedra sheets is again reflected from the broadening of the peaks and the merge of the second and the third peaks in the pair-distribution functions (Fig. 5 (a)), the splits of the peak around $60^0$ degrees in the angle-distribution functions (see Fig. 5 (b)), and the broadening of peak around $110^0$ degrees in the angle-distribution functions (Fig. 5 (b)).

The complex chemical bonding natures in these low-dimensional icosahedra structures alternate their electronic properties as compared to other boron allotropes (e.g., the gapless materials of the monolayer buckled α and $δ_6$ sheets, and the flat $δ_4$-sheets, and the semiconductor materials of the α-, β-, and γ-B ). Fig. 6 (a)-(d) shows the electronic densities of states (DOS) of these low-dimensional icosahedra structures together with that of the α-B. The Fermi energy is presented by the red dashed line at zero. Obtained energy gap for the α-B (Fig. 6 (e)) is 1.90 eV which is only about 0.1 eV underestimate compared to the optical gap of 2.0 eV measured for the α-B [51]. Calculated energy gap for the icosahedra chain, on the other hand, is 0.74 eV, which is smaller than that of the α-B but still maintains the semi-conducting behavior. Obtained energy gap for the icosahedra sheet, however, strongly depends on the symmetry of the sheet. The icosahedra $δ_6$ and $δ_4$ sheets behave semiconducting nature (e.g., 0.52 eV for the icosahedra $δ_6$ sheet and 0.39 eV for the icosahedra $δ_4$ sheet, respectively). While, the icosahedra α is found to be a gapless material (~0.018 eV). We should note that since the calculated the energy gap for the α-B is underestimated by 0.1 eV, the calculated energy gaps for the low-dimensional icosahedra structures are expected not to have significant underestimate and the conclusion that the icosahedra chain and the icosahedra $δ_6$ and $δ_4$ sheets are semiconductor materials is acceptable.

One more interesting issue from the synthesis point of view is to find the possible pathways to form the icosahedra α and $δ_4$ sheets from the icosahedra $δ_6$ sheet and how much the external energy is required to transfer them from one phase to another along the pathways. To shed light into the possible pathways (e.g., from the $δ_6$ phase to α/$δ_4$ and from α to $δ_4$ phases), we modeled such phase transition process by gradually removing the icosahedrons out of the sheet plane. The transition from the icosahedra $δ_6$ phase to the icosahedra α/$δ_4$ phase is molded by gradually removing four/nine icosahedrons from the 6x6 unit cell of the $δ_6$ phase along the direction perpendicular to the icosahedra $δ_6$ sheet (see Fig. 7 (a) and (b)). The transition from the α phase to the $δ_4$ phase is molded by gradually removing five icosahedrons from the 2x2 unit cell of the α phase along the direction perpendicular to the sheet and moving another three icosahedrons to the hexagonal holes in the sheet plane (see Fig. 7 (c), respectively. The corresponding energy barriers of such transitions are then estimated from the energy difference between the initial stage (e.g., the icosahedra $\delta_6$ or $\alpha$ phase) and the final stage (e.g., the combined system of the icosahedra $\alpha$ sheet plus four isolated icosahedra $B_{12}$, or the icosahedra $\delta_4$ sheet plus nine isolated icosahedra $B_{12}$, or the icosahedra $\delta_4$ sheet plus five isolated icosahedra $B_{12}$, respectively). The results are presented in Fig. 7 (d). We found that it will require 0.17eV/atom (~ 1700 K) to transfer the icosahedra $\delta_6$ phase to the icosahedra $\alpha$ phase, and 0.27eV/atom (~ 2700 K) to transfer the icosahedra $\alpha$ phase to the icosahedra $\delta_4$ phase. But it will require much larger energy 0.38 eV/atom (~ 3800 K) to transfer the $\delta_6$ phase to the $\delta_4$ phase. These high energy barriers indicate that the transition among these three phases is not easy to happen. In another word, these icosahedra sheets are relatively stable and are possible to be synthesized.

## IV. CONCLUSION

Four new boron allotropic structures have been found. They are the icosahedra chain, the icosahedra $\delta_6$ sheet, the icosahedra $\alpha$ sheet, and the icosahedra $\delta_4$ sheet, respectively. These novel icosahedra structures have shown to be structurally and energetically stable. The icosahedra $\delta_6$ sheet is energetically most stable among these novel icosahedra structures, followed by the icosahedra $\alpha$ sheet, the icosahedra $\delta_4$ sheet, and then the icosahedra chain. The slight deformation of the icosahedra $B_{12}$ in these new icosahedra structures due to the existence of the dangling bonds leads to the broad distributions in the bond length and the angles between boron atoms inside the icosahedron. All the inter-icosahedra bonds show two-center directional covalent bonding nature. Especially, different from the three-center type of bonding nature in the $\alpha$-B, the $\delta$ bonds in the icosahedra sheets have the two-center bonding nature and always form as a pair to the neighboring icosahedrons. The short bond length of such bonds, as compared to $\alpha$-B, also indicates the strong interaction between the NN icosahedra $B_{12}$ in the icosahedra sheet structures. Furthermore, the distribution of the inter-icosahedra δ bonds in the icosahedra sheets depends on the environments of the icosahedra $B_{12}$. More interesting finding is that these new icosahedra structures behave like semiconductor materials except the icosahedra α sheet which behaves like a gapless material. The high energy barriers between the stable icosahedra sheets indicate their relative stability and possible transition pathways could provide the information for experiment synthesis.

Acknowledgements

The first author, C. B. Kah, acknowledges the strong support for his Ph. D. thesis by the McSweeny Fellowship awarded through the College of Arts and Sciences of the University of Louisville.

References

[1] Murat Atiş, Cem Özdoğan, and Ziya B. Güvenç, Int. J. of Quantum Chemistry **107**, 729 (2007)

[2] I. Boustani, Phys. Rev. B **55**, 16426 (1997)

[3] I. Boustani, Chemical Physics Letters **240**, 135 (1995)

[4] I. Boustani, Chemical Physics Letters **233**, 273 (1995)

[5] I. Boustani, Surf. Sci. **370**, 355 (1997)

[6] I. Boustani, J. Solid State Chemistry **133**, 182 (1997)

[7] Constantin Romanescu, Dan J. Harding, André Fielicke, and Lai-Sheng Wang, J. Chem. Phys. **137**, 014317 (2012)

[8] I. Boustani, Zhen Zhu, and David Tománek, Phys. Rev. B **83**, 193405 (2011)

[9] Zachary A. Piazza, Han-Shi Hi, Wei-Li Li, Ya-Fan Zhao, Jun Li, and Lai-Sheng Wang, Nature Communication, DOI: 20.1038/ncomms4113 (2014)

[10] Hua-Jin Zhai, Ya-Fan Zhao, Wei-Li Li, Qian Chen, Hui Bai, Han-Shi Hu, Zachary A. Piazza, Wen-Juan Tian, Hai-Gang Lu, Yan-Bo Wu, Yue-Wen Mu, Guang-Feng Wei, Zhi-Pan Liu, Jun Li, Si-Dian Li, and Lai-Sheng Wang, Nat. Chem. DOI: 10.1038/NCHM.1999 (2014)

[11] Nevill Gonzalez Szwacki, Nanoscale Research Letters **3**, 49 (2008)

[12] N. Gonzalez Szwacki, A. Sadrzadeh, and B.I. Yakobson, Phys. Rev. Letts. **98**, 166804 (2007)

[13] G. Gopakumar, M.T. Nguyen, and A. Ceulemans, Chem. Phys. Letts. **450**, 175 (2008)

[14] A. Ceulemans, J. Tshishimbi, G. Gopakumar, and M. T. Nguyen, Chem. Phys. Letts. **461**, 226 (2008)

[15] Tunna Baruah, Mark R. Pederson, and Rajendra R. Zope, Phys. Rev. B **78**, 045408 (2008)

[16] Rosi N. Gunasinghe, Cherno B. Kah, Kregg D. Quarles, and Xiao-Qian Wang, Appl. Phys. Letts. **98**, 261906 (2011)

[17] N. Gonzalez Szwacki, A. Sadrzadeh, and B.I. Yakobson, Phys. Rev. Letts. **100**, 159901 (2008)

[18] P. Tandy, M. Yu, C. Leahy, C.S. Jayanthi, and S.Y. Wu, arXiv:1408.4931 [cond-mat.mtrl-sci] (2014)

[19] Dasari L. V. K. Prasad and Eluvathingal D. Jemmis, Phys. Rev. Letts. **100**, 165504 (2008)

[20] Rajendra R. Zope, EPL **85**, 68005 (2009)

[21] Ihsan Boustani, Alexander Quandt, Eduardo Hernández, and Angel Rubio, J. Chem. Phys. **110**, 3176 (1999)

[22] Abhishek K. Singh, Arta Sadrzadeh, and Boris I. Yakobson, Nano Letts. **8**, 1314 (2008)

[23] Xiaobao Yang, Yi Ding, and Jun Ni, Phys. Rev. B **77**, 041402(R) (2008)

[24] Hui Tang and Sohrab Isnail-Beigi, Phys. Rev. B **82**, 115412 (2010)

[25] Hui Tang and Sohrab Isnail-Beigi, Phys. Rev. Letts. **99**, 115501 (2007)

[26] Xiaojun Wu, Jun Dai, Yu Zhao, Zhiwen Zhou, Jinlong Yang, and Xiao Cheng Zeng, ACS NANO **6**, 7443 (2012)

[27] David Emin, Physics Today **40**, 55 (1987)

[28] Artem. R. Oganov, Jiuhua. Chen, Carlo. Gatti, Yanzhang. Ma, Yanming Ma, Colin W. Glass, Zhengxian Liu, Tony Yu, Oleksandr O. Kurakevych, and Vladimir L. Solozhenko, Nature **457**, 863 (2009)

[29] Gleb Parakonsky, Natalia Dubrovinskaia, Elena Bykova, Richard Wirth and Leonid Dubrovinsky, Scientific Reports 1: 96. Published online 2011 September 19. doi: 10.1038/srep00096

[30] Tadashi Ogitsu, François Gygi, John Reed, Yukitoshi Motome, Eric Schwegler, and Giulia Galli, J. AM. Chem. Soc. **131**, 1903 (2009)

[31] Chao Jiang, Zhijun Lin, Jianzhong Zhang, and Yusheng Zhao, Appl. Phys. Letts. **94**, 191906 (2009)

[32] B. F. Decker and J. S. Kasper, Acta Crystallography. **12**, 503 (1959)

[33] A. Masago, K. Shirai, and H. Katayama-Yoshida, Phys. Rev. B **73**, 104102 (2006)

[34] Carolyn Jones Otten, Oleg R. Lourie, Min-Feng Yu, John M. Cowley, Mark J. Dyer, Rodney S. Ruoff, and William E. Buhro, J. AM. Chem. Soc. **124**, 4564 (2002)

[35] X.M. Meng, J.Q. Hu, Y. Jiang, C.S. Lee, S.T. Lee, Chem. Phys. Letts. **370**, 825 (2003)

[36] Yingjiu Zhang, Hiroki Ago, Motto Yumura, Toshiki Komatsu, Satoshi Ohshima, Kunio Uchida, and Sumio Iijima, Chem. Commun., 2806 (2002)

[37] C. Leahy, M. Yu, C. S. Jayanthi, and S. Y. Wu, Phys. Rev. B **74**, 155408 (2006)

[38] M. Yu, S. Y. Wu and C. S. Jayanthi, Physica E **42**, 1 (2009)

[39] Ming Yu, Indira Chaudhuri, C. Leahy, C.S. Jayanthi, and S.Y. Wu, J. Chem. Phys. **130**, 184708 (2009)

[40] W.Q. Tian, M. Yu, C. Leahy, C.S. Jayanthi, and S.-Y. Wu, J. of Computational and Theoretical Nanoscience **6**, 390 (2009)

[41] M. Yu, C.S. Jayanthi, and S.Y. Wu, Nanotechnology **23**, 235705 (2012)

[42] M. Yu, C. S. Jayanthi, and S. Y. Wu, J. Material Research **28**, 57 (2013)

[43] Z. H. Xin, C. Y. Zhang, M. Yu, C. S. Jayanthi, and S. Y. Wu, Computational Materials Science **84**, 49 (2014)

[44] I. Chaudhuri , Ming Yu, C. S. Jayanthi, and S. Y. Wu, J. Phys. Condensed Matter **26**, 115301 (2014)

[45] Xiaojun Wu, Jun Dai, Yu Zhao, Zhiwen Zhou, Jinlong Yang, and Xiao Cheng Zeng, ACS Nano **6**, 7443 (2012)

[46] G. Kresse and J. Hafner, Phys. Rev. B **48**, 13115 (1993)

[47] D. Vanderbilt, Phys. Rev. B **41**, 7892 (1990)

[48] G. Kresse and D. Joubert, Phys. Rev. B **59**, 1758 (1999)

[49] J. P. Perdew, in *Electronic Structure of Solids '91,* edited by P. Ziesche and H. Eschrig (Akademie Verlag, Berlin,1991), p. 11.

[50] H.C. Longuet-Higgins and M. de V. Roberts, Proc. R. Soc. London, Ser. A **230**, 110 (1955)

[51] O.A. Golikova, N.E. Solovev, Ya. A. Ugai, and V. A. Feigelman, Phys. Stat. Sol. **86**, K51 (1984)

Table 1 Comparison of the structural properties of the icosahedra chain and sheets with the icosahedra $B_{12}$ ball and α-B. The 1$^{st}$ column: the system notation; the 2$^{nd}$ column: the bond lengths between boron atoms inside the icosahedra $B_{12}$, $b_{intra}$; the 3$^{rd}$ column: the bond lengths of the inter-icosahedra bonds, $b_{inter}$, where S and D in the parentheses denote the two types of the inter-icosahedra bonds (see their definitions in the text); the 4$^{th}$ column: the number of the single inter-icosahedra bonds per icosahedra $B_{12}$, $N_{inter}^{S}$; the 5$^{th}$ column: the number of the pairs of the inter-icosahedra δ bonds per icosahedra $B_{12}$, $N_{inter}^{D}$.

| System | | $b_{intra}$ (Å) | $b_{inter}$ (Å) | $N_{inter}^{S}$ | $N_{inter}^{D}$ |
|---|---|---|---|---|---|
| Icosahedra $B_{12}$ | | 1.61-1.75 | NA | NA | NA |
| α-B | | 1.73-1.80 | 1.67(S); 2.03 (D) | 6 | 0 |
| Icosahedra chain | | 1.62-1.75 | 1.76 (S) | 2 | 0 |
| Icosahedra sheets | $δ_6$ | 1.56-1.77 | 1.69 (S); 1.80 (D) | 4 | 2 |
| | α | 1.56-1.77 (Type A) <br> 1.57-1.80 (Type B) <br> 1.59-1.79 (Type C) | 1.69 (S); 1.82 (D) (Type A) <br> 1.69-1.72 (S); 1.82 (D) (Type B) <br> 1.69-1.72 (S); 1.75 (D) (Type C) | 4 (Type A) <br> 3 (Type B) <br> 4 (Type C) | 2 (Type A) <br> 2 (Type B) <br> 1 (Type C) |
| | $δ_4$ | 1.59-1.75 (Type A) <br> 1.59-1.80 (Type B) | 1.68 (S) (Type A) <br> 1.68 (S); 1.82 (D) (Type B) | 4 (Type A) <br> 2 (Type B) | 0 (Type A) <br> 2 (Type B) |

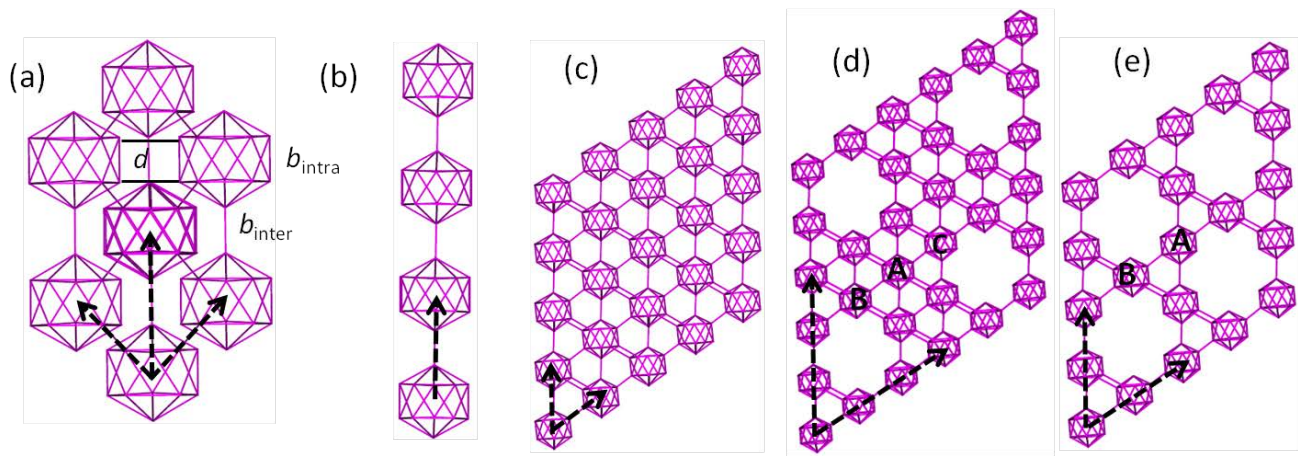

Figure 1

Figure 1 The schematic illustration of the structures for (a) the α-B, (b) the icosahedra chain, (c) the icosahedra $\delta_6$ sheet, (d) the icosahedra α sheet, and (e) the icosahedra $\delta_4$ sheet, respectively. The lattice vectors for each structure are denoted by the dashed arrows. The notations $b_{intra}$, $b_{inter}$, and $d$ in (a) indicate the bonds between boron atoms inside the icosahedra $B_{12}$, the bonds between the nearest neighbor icosahedra $B_{12}$ (referred as the single inter-icosahedra bond), and the distance between the next nearest neighbor icosahedra $B_{12}$ (referred as the inter-icosahedra δ bond), respectively. Note that these parallel δ bonds in the α-B and the icosahedra sheets are perpendicular to the edges of the icosahedrons connected to the bonds. The capital letters in (d) and (e) indicate the different types of inter-icosahedra bonding nature (see the definitions in the context) for each icosahedra $B_{12}$.

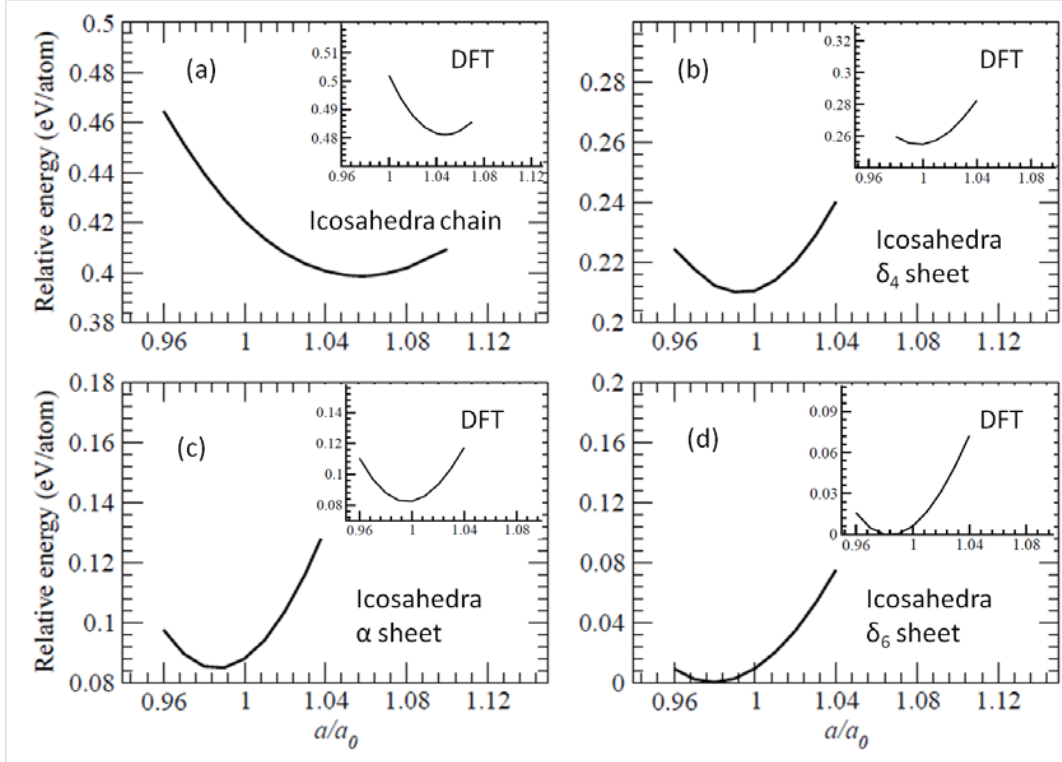

Figure 2

Figure 2 The relative energy per atom (see the definition in the context) versus the ratio of the lattice constant $a$ to that of the α-B at the equilibrium $a_0$ for (a) the icosahedra chain, (b) the $\delta_4$ icosahedra sheet, (c) the α icosahedra sheet, and (d) the $\delta_6$ icosahedra sheet, respectively. The corresponding DFT results are presented in the insets.

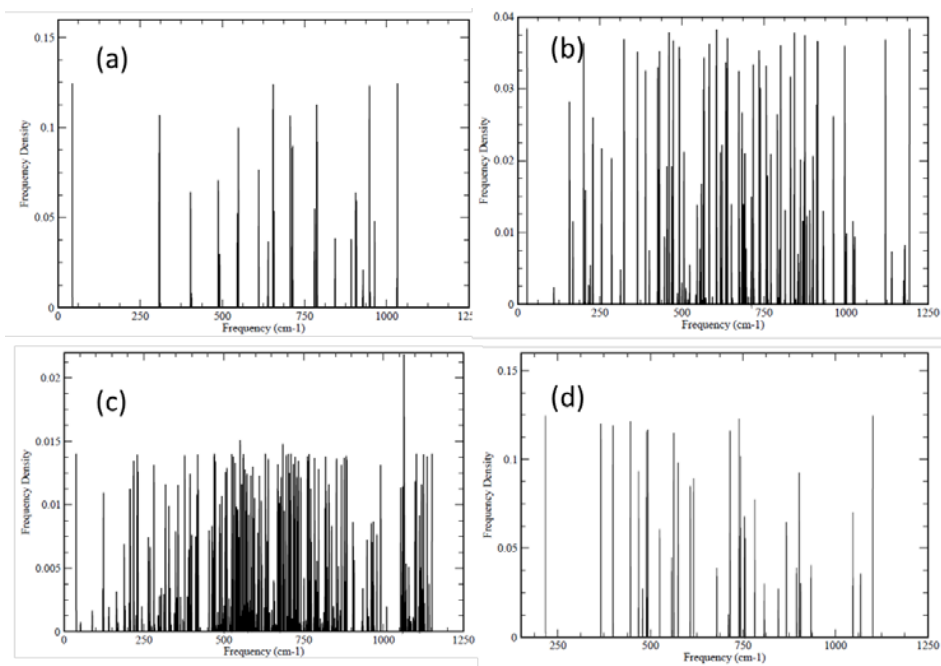

Fig. 3

Figure 3 The vibrational frequency densities of (a) the icosahedra chain, (b) the icosahedra $\delta_4$ sheet, (c) the icosahedra α sheet, and (d) the icosahedra $\delta_6$ sheet, respectively.

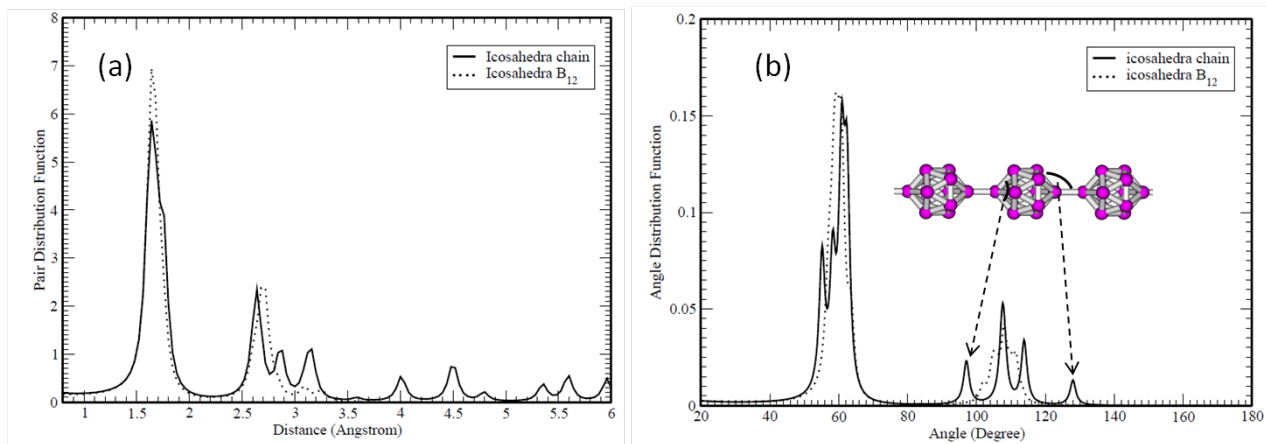

Figure 4

Figure 4 (a) The pair-distribution functions of the icosahedra chain (black solid curve) and the icosahedra $B_{12}$ (black dashed curve). (b) The angle-distribution functions of the icosahedra chain (black solid curve) and the icosahedra $B_{12}$ (black dashed curve). The inset in (b) is the structure of the icosahedra chain. The peaks around $96^0$ and $128^0$ degrees in the angel-distribution function of the icosahedra chain are indicated by the dashed arrows.

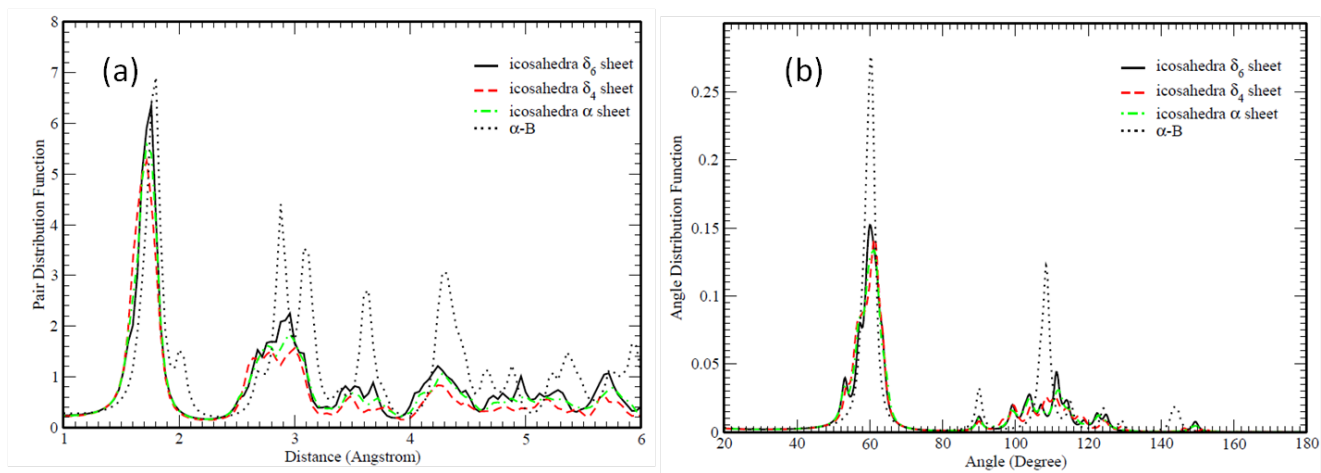

Figure 5

Figure 5 (a) The pair-distribution functions of the icosahedra $\delta_6$ sheet (black solid curve), the icosahedra $\delta_4$ sheet (red dashed curve), the icosahedra α sheet (green dotted-dash curve), and the α-B (black dotted curve), respectively. (b) The pair-distribution functions of the icosahedra $\delta_6$ sheet (black solid curve), the icosahedra $\delta_4$ sheet (red dashed curve), the icosahedra α sheet (green dotted-dash curve), and the α-B (black dotted curve), respectively. Note that the peak around $90^0$ degree in (b) represents the two-parallel inter-icosahedra δ bonds which are perpendicular to the edges of the icosahedra $B_{12}$ (see Fig. 1 (a), (c)-(e)).

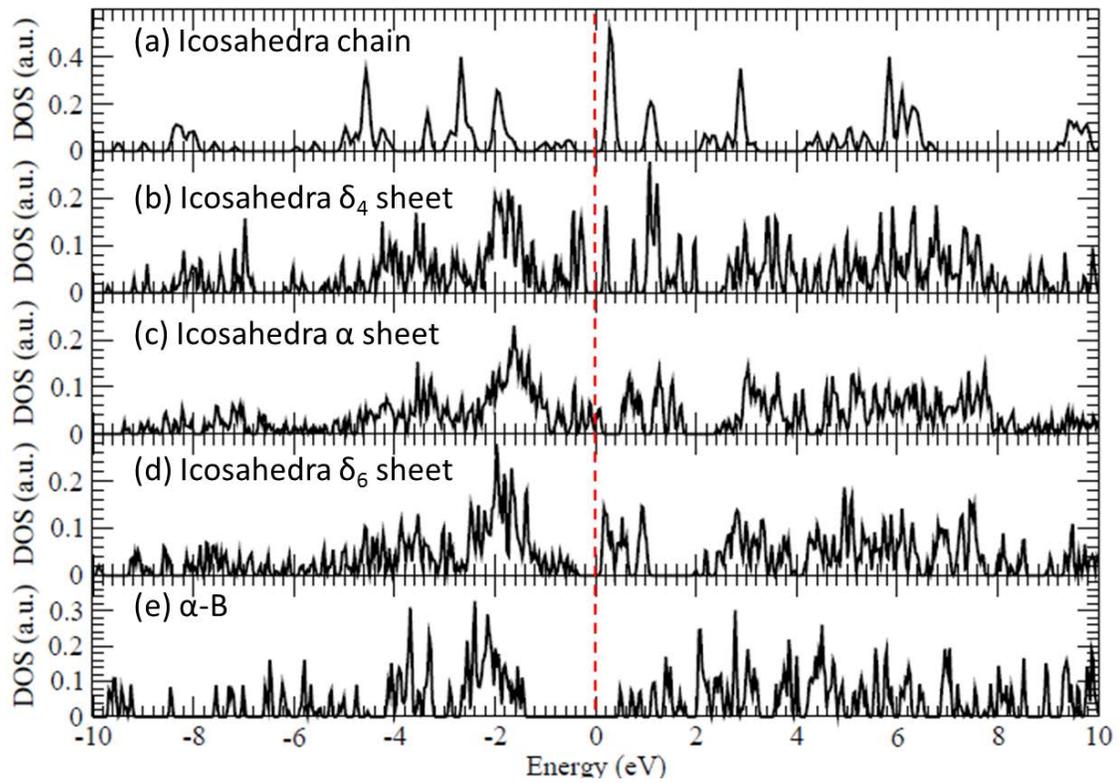

Figure 6

Figure 6 The densities of states (DOS) of the icosahedra chain (a), the icosahedra $\delta_4$ sheet (b), the icosahedra α sheet (c), the icosahedra $\delta_6$ sheet (d), and the α-B (e), respectively. The Fermi energy is presented by the red dashed vertical line.

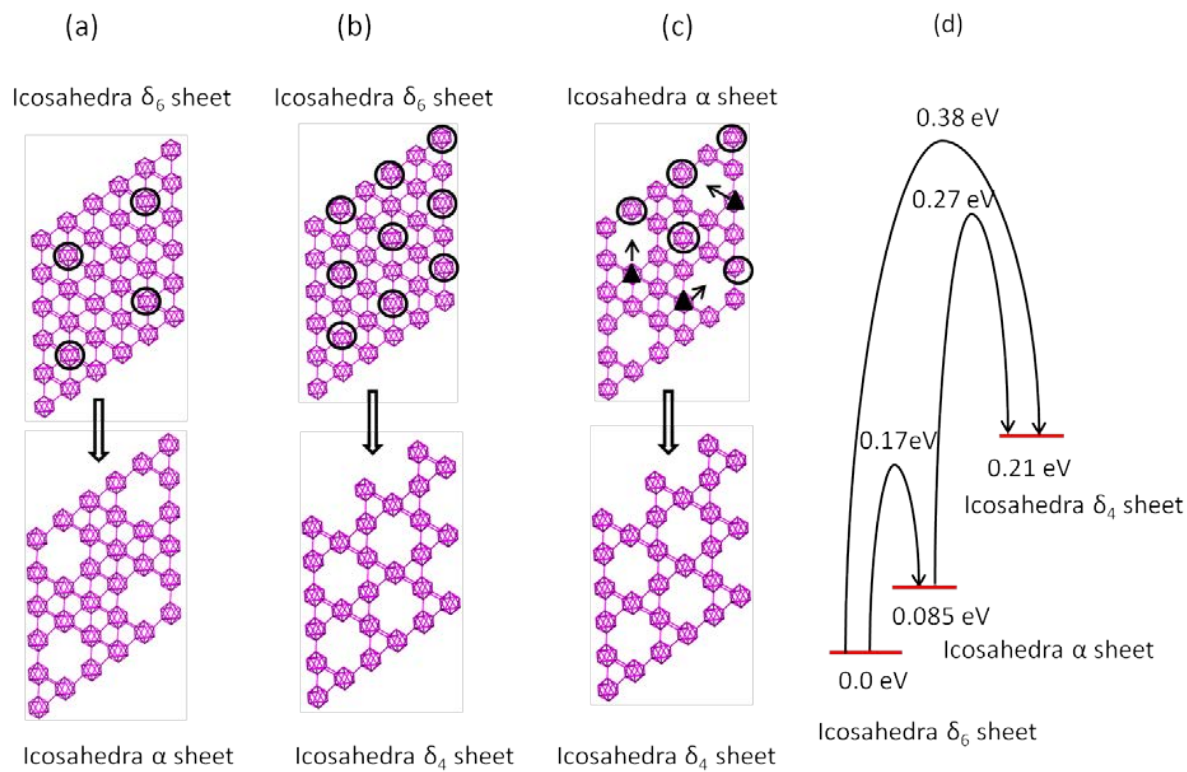

Figure 7

Figure 7 The schematic illustration of (a) the transition from the icosahedra $\delta_6$ sheet (up) to the icosahedra $\alpha$ sheet (down), (b) the transition from the icosahedra $\delta_6$ sheet (up) to the icosahedra $\delta_4$ sheet (down), (c) the transition from the icosahedra $\alpha$ sheet (up) to the icosahedra $\delta_4$ sheet (down), and (d) the energy barriers (numbers associated at peak of the black curves) of the transition from the icosahedra $\delta_6$ sheet to the icosahedra $\alpha/\delta_4$ sheet and from the icosahedra $\alpha$ sheet to the icosahedra $\delta_4$ sheet, respectively. The black circles in (a)-(c) denote the removed icosahedrons, and the black triangles in (c), denote the icosahedrons moved to the center of the hexagons indicated by the arrows. The numbers associated with the red lines in (d) are the relative cohesive energy per atom of the optimized icosahedra sheets to that of the icosahedra $\delta_6$ sheet.